# TECVis: A Visual Analytics Tool to Compare People's Emotion Feelings

Ilya Nemtsov,* MST Jasmine Jahan,† Chuting Yan,‡ and Shah Rukh Humayoun§

Department of Computer Science, San Francisco State University, USA

**ABSTRACT**

Twitter is one of the popular social media platforms where people share news or reactions towards an event or topic using short text messages called "tweets". Emotion analysis in these tweets can play a vital role in understanding peoples' feelings towards the underlying event or topic. In this work, we present our visual analytics tool, called TECVis, that focuses on providing comparison views of peoples' emotion feelings in tweets towards an event or topic. The comparison is done based on geolocations or timestamps. TECVis provides several interaction and filtering options for navigation and better exploration of underlying tweet data for emotion feelings comparison.

## 1 INTRODUCTION

The rise of social media platforms (e.g., Facebook, Twitter, Instagram, etc.) has transformed the way we access and disseminate information. They allow us to obtain not only news and updates from multiple sources but provide a powerful medium for the people to express their feelings and reactions towards events or topics. Among these platforms, Twitter (recently renamed as X), a microblogging service, has become one of the most influential platforms, with over 450 million monthly active users as of 2023 [12]. The platform enables users to share their feelings and news through short text messages called "tweets". These tweets often come with additional metadata (e.g., user profile, geolocation, timestamp, etc.) creating a valuable information resource for various analytical purposes.

This extensive pool of data at Twitter provides unique opportunities for researchers, businesses, and policymakers to uncover patterns, trends, and insights that can then be applied to diverse fields such as marketing, political science, public health, disaster management, etc. Targeting this, many researchers have developed visualization tools to explore and understand peoples' tweets from different perspectives, e.g., showing the sentimental analysis of tweets (e.g., [3] [4] [9]), analyzing and exploring tweet data based on the geo-spatial information (e.g., [2] [10] [6]), or exploring topics and analyzing their evolution over time (e.g., [1] [5] [13]). While some other researchers provided new angle to analyze tweet data, e.g., detecting unexpected events [7], building impression of a user [8], stance exploration in social media [11], and exploring relationships between the frequent keywords [4].

One way to better explore tweet data is about comparison of peoples' reaction towards the same event from different geolocations, e.g., peoples' reaction towards COVID-19 or their views regarding recent US Supreme Court ruling in 2022 to overturn Roe v. Wade abortion right. Such comparison exploration of tweet data, which is missing in previous work, would help us analyzing better public views towards some events or topics, which could be useful to understand better the political and social norms of different geolocations.

Targeting this concern, we developed a visual analytics tool, called **TECVis** (**T**weets' **E**motions **C**omparison **Vis**ualizer), that focuses on comparison of peoples' feelings towards an event or topic using the emotions in tweets (i.e.: *anger*, *fear*, *anticipation*, *trust*, *surprise*, *sadness*, *joy*, and *disgust*) and sentiment analysis (i.e.: *negative*, *neutral*, and *positive*). The comparison is done based on geolocations or timestamps, where users would be able to explore peoples' feelings towards the same event or topic from different geolocations or from one timestamp to another timestamp. TECVis provides several interaction and filtering options so users can navigate the underlying tweet data for better exploration.

## 2 THE DATASET

As a proof of concept, we targeted the COVID-19 topic and collected the dataset through the Twitter API[1], using the COVID-19 and related keywords from the time period January 2021 to May 2021 based on US geolocations. The final dataset consisted of 574,806 tweets (only English tweets and US geolocations) from an initial 952,278 collected tweets. We used the Natural Language Toolkit (NLTK[2]) Python library to tokenize the keywords. We used VADER[3] (Valence Aware Dictionary and sEntiment Reasoner) library to get sentiment polarities of tweets and their confidence scores. Emotion analysis plays a vital role in understanding an individual's feelings and emotions when analyzing textual data. We used NRC Emotion Lexicon library[4] to get emotions for each tweet. The library provides eight basic emotions, as mentioned earlier, to the given text with association scores ranging from 0 to 1 and two sentiments (i.e., negative and positive).

## 3 TECVIS: TWEETS' EMOTIONS COMPARISON VISUALIZER

We developed our TECVis tool as a web-based tool where the client-side was developed using HTML, CSS, and JavaScript. We used React framework for managing the frontend development and D3.js library for providing the resulting visualizations, while the server side was developed using the Node.js.

The TECVis tool provides the comparison of peoples' feelings towards an event or topic using the emotions and sentiment polarities in tweets based on two settings. In the first setting, the comparison is based on geolocations, which is useful to explore peoples' feelings and reactions from different locations towards the same event. While in the second setting, TECVis provides the comparison based on timestamps, which is useful to explore the changes in peoples' feelings and reactions towards the same event over a period of time.

Figure 1 shows the overview of the TECVis tool based on geolocations comparison. For showing the eight feelings in emotion comparison, we use a dot plot (see Fig. 1), where the x-axis is used to display the average score ranging from 0 to 1 for each feeling. On the y-axis, all the geolocations (i.e., US states in the current dataset) are listed. On the dot-plot, each dot represents the average score of each emotion feeling of all the associated tweets from the underlying geolocation. We use different colors for each emotion feeling.

---

*e-mail: nemtsovilya@gmail.com
†e-mail: jammy.sust@gmail.com
‡e-mail: christinayan1019@gmail.com
§e-mail: humayoun@sfsu.edu

[1]https://developer.twitter.com/en/docs/twitter-api
[2]https://www.nltk.org/
[3]https://github.com/cjhutto/vaderSentiment
[4]https://pypi.org/project/NRCLex/

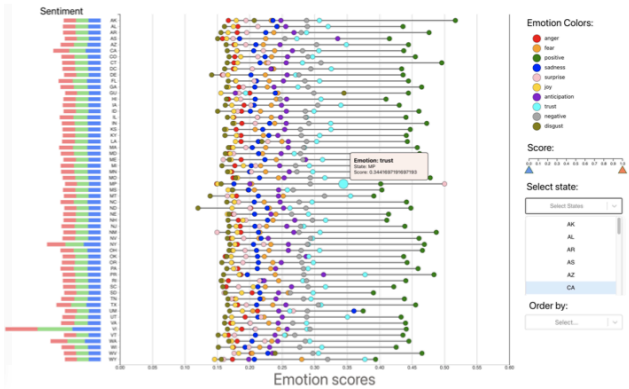

Figure 1: TECVis overview with COVID-19 dataset. The middle dot plots show eight emotion feelings' average scores associated to each geolocation, a US state in this view. The left-side horizontal bars, associated with each geolocation, show tweets' count and their sentiment polarities. The right-side panel provides filtering options.

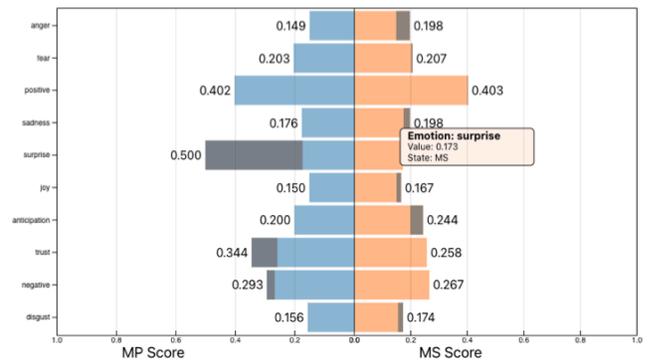

Figure 2: TECVis uses a Tornado chart for the side-by-side comparison of two geolocations or timestamps.

We categorize emotion feelings into positive feelings category (i.e.: *anticipation*, *trust*, *surprise*, and *joy*) and negative feelings category (i.e.: *anger*, *fear*, *sadness*, and *disgust*). We noticed that the NRC Emotion Lexicon library may provide score in both kind of feelings, as sometimes it is difficult to extract the exact feelings based on the short text in tweets. Therefore, we decided to associate a tweet either to positive feelings category or negative feelings category based on which category has higher aggregated value. Furthermore, for a better mean value we consider a feeling towards a tweet only if it has a value above the 0.1 threshold.

On the left side of each geolocation, TECVis uses a horizontal bar to represent the associated tweets' count to this geolocation. Each bar also shows the associated sentiment polarity distribution (i.e., red for negative, blue for neutral, and green for positive). Each polarity color length represents the count of associated tweets. Mouse hover a particular bar provides further details through a tooltip. The user can switch to the timestamp view, where y-axis is then used for comparison based on timestamp (e.g., days, weeks, or months).

For a side-by-side comparison of two geolocations or timestamps (users can click to select these on the main view), TECVis shows a new pop-up view with a Tornado chart (see Fig. 2. In this Tornado chart, each side shows emotion feeling scores of one selected geolocation or timestamp. For providing the differences between the scores, TECVis highlights the difference value using a darker color in the higher score side. This gives a quick indication of not only the high score value of an emotion feeling from one geolocation/timestamp but also the level of difference between them.

TECVis provides several interaction, filtering, and navigation options: Users can navigate from geolocation comparison to timestamp comparison and vice versa. For example, when users select a particular geolocation then TECVis updates the current view with showing the underlying geolocation tweet data with comparison perspective of timestamps, where users can see the comparison by days, weeks, or months. This option also works from the main timestamp comparison view to geolocation comparison view. TECVis also provides the facility to filter the data based on selected geolocations (see right-side panel in Fig. 1) or timestamps . Users can also filter the data based on a particular emotion feeling or the emotion feeling score range using a score range bar (see right-side panel in Fig. 1).

## 4 CONCLUDING REMARKS

Visual comparison and exploration of peoples' feelings, based on geolocations or timestamps, towards an event or topic using the emotions and sentiment polarities in tweets could be useful for understanding better the political and social norms of different geolocations towasrds the same event or topic. In the future, we plan to conduct detailed user study to evaluate the tool from the common usability aspects as well as explorative user study to find out how analysts can explore and compare political or social norms of different geolocations using the peoples' emotion feelings in tweets using different datasets. We also intend to open source the tool so researchers and analysts can use their datasets for exploring different events or topics.